\documentclass{ws-ijmpe}
\usepackage{graphicx}

\begin{document}

\title{NUCLEAR PERIPHERY IN MEAN-FIELD MODELS}
\author{A.~BARAN and P.~MIERZY\'NSKI}
\address{Institute of Physics, University of M. Curie-Sk\l odowska, \\ 
ul. Radziszewskiego 10, 20-031 Lublin, Poland}
\markboth{A. Baran and P. Mierzy\'nski}{Nuclear periphery in mean field models}

\maketitle
\begin{history}
\received{(October 12, 2003)}
\revised{(revised date)}
\end{history}

\begin{abstract}
The halo factor is one of the experimental data which describes 
a distribution of neutrons in nuclear periphery. 
In the presented paper we use Skyrme-Hartree (SH)
and the Relativistic Mean Field (RMF) models and we calculate
the neutron excess factor $\Delta_B$ defined in the paper which 
differs slightly from halo factor $f_{\rm exp}$.
The results of the calculations are compared to the measured data.
\end{abstract}

\section{Introduction.}

One of the important problems of nuclear physics is the description  of
the periphery of the nucleus.
From theoretical and experimental point 
of view the topic is complex
and much information is needed in order 
to predict neutron or proton excess
in the neighbourhood of nuclear 
surface.\cite{Lubinski98,Schmidt99,Baran97} 
The peripheral density distributions of protons and neutrons
are determined theoretically by single particle eigenfunctions 
of protons and neutrons in this region. Applying different 
nuclear models one calculates wave functions in the deformed 
oscillator basis. They behave asymptotically like 
$\exp(-\gamma r^2)$, where $\gamma$ is a constant and $r$ is 
the distance from 
the center of the nucleus and do not reproduce properly
the particle densities at large distances. The real nuclear 
densities behave like $\exp(-\beta r)$.
In the following we rather stay in the nuclear interior where
the density distributions are well defined
and we describe the nuclear peripheral properties using only
well defined theoretical internal nuclear densities. 
We would like to show that a simple 
function $\delta_B(\vec r)$ introduced by Bethe\cite{Bethe70} 
\begin{eqnarray}
\delta_{B}(\vec{r})={Z \over N}\rho_{N}(\vec{r})-\rho_{Z}(\vec{r})=
            \tilde{\rho}_{N}(\vec{r})-\rho_{Z}(\vec{r})\,,
\label{eq-deltab}
\end{eqnarray}
allows to predict the experimentally observed halo data
without the knowledge of the local behaviour of a tails of 
nuclear densities in asymptotic region.
The function given by Eq.(\ref{eq-deltab}) is the 
difference of the relative neutron density 
$\tilde{\rho}_{N}(\vec{r})={(Z/N)}\rho_N$ and the proton 
density $\rho_{Z}(\vec{r})$ 
at a point $\vec r$. 

\begin{figure}
\begin{center}
\includegraphics[scale=0.45]{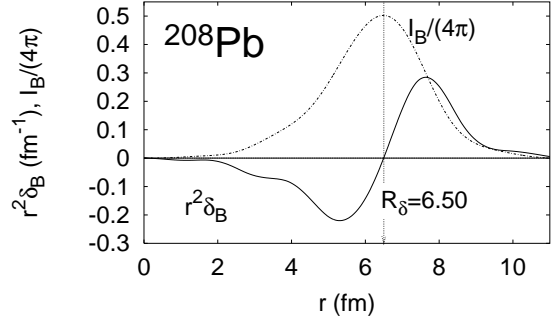}
\caption{Bethe function of $^{208}$Pb (multiplied by $r^2$, 
solid line) and the integral $I_B(r)$ (dashed line). The maximum of 
$I_B$ corresponding to $R=6.5$fm (zero of $\delta_B(r)$) 
is the neutron excess factor discussed in the text. \label{fig-dpb}}
\end{center}
\end{figure}

In the following section we introduce a neutron excess factor 
(NEF) based on $\delta_B$ defined in Eq. (\ref{eq-deltab}) function 
and we  calculate NEF for a class of nuclei studied at Low Energy 
Antiproton Ring (LEAR) 
facility in CERN.\cite{Lubinski98,Schmidt99}. Two mean field models are 
used in calculating NEF which after that is compared to the experimental
data.

\section{Model}
It is a trivial to notice that the integral of the Bethe 
function (\ref{eq-deltab}) over the whole space vanishes
\begin{equation}
\int\delta_{B}(\vec{r})\,d^{3}\vec{r}=0\,.
\label{eq-dint}
\end{equation}
This direct property allows to define the total peripheral neutron or
proton excess without the knowledge of the density distribution
in the nuclear periphery. For example, in the case of the spherical symmetry
equation (\ref{eq-dint}) reads
\begin{equation}
4\pi\int_0^\infty r^2 \delta_B(r)\,dr = 
4\pi\int_0^{r'} r^2 \delta_B(r)\,dr+
4\pi\int_{r'}^\infty r^2 \delta_B(r)\,dr = 0\,,
\end{equation}
where $r'$ is an arbitrary point which if chosen properly
will maximize the first term in the above
equation. Let us define the following function of the distance $r$
\begin{equation}
I_{B}(r)=-4\pi\int_{0}^r r'^2\,dr'\,.
\label{eq-i}
\end{equation}
The maximal value of $\Delta_B$ with respect to $r$
\begin{equation}
\Delta_{B}= {\rm max}_{r} I_B(r)\,,
\label{eq-nef}
\end{equation}
gives the relative outer excess of neutrons or protons out of the 
point which maximizes $I_B$. The positive value of $\Delta_B$ 
indicates the excess of neutrons while
the negative value of it corresponds to the case of the relative 
proton excess.  In the following we call the value of $\Delta_B$ 
the neutron excess factor (NEF).
Both $\delta_B$ and $_B$ functions for $^{208}$Pb are shown 
in Figure \ref{fig-dpb}.

Similar functions and quantities are defined for more general
case of a deformed nuclei.\cite{Baran04} 
The corresponding integrals and quantities (Eqs. \ref{eq-i} and \ref{eq-nef})
are calculated in this case by
taking the integrals over the region of the deformed surface similar 
to the average nuclear surface.

\section{Antiproton annihilation on nucleus}

The measurements of products of the annihilation of antiprotons
on nuclei\cite{Lubinski98,Schmidt99} inform on the value of the neutron excess in
the nuclear periphery where the annihilation takes place. 
This is the outer region extending over the whole 
periphery of the nucleus. From the experiment\cite{Lubinski98,Schmidt99} 
one approximatelly knows the average
radius of stopping of antiprotons. It is slightly larger (by about 2fm)
than the average nuclear radius $R=r_0A^{1/3}$.

\begin{figure}
\begin{center}
\includegraphics[scale=3.5]{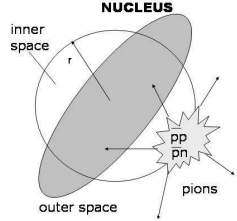}
\end{center}
\caption{
Annihilation of $\bar pp$ or $\bar pn\;$ 
on the periphery of the deformed nucleus. The inner and the outer 
regions are divided by a sphere of radius $r$. The pions are the 
outcome of the annihilation reaction.\label{fig-annih}}
\end{figure}

While the deformed nuclei are not oriented in the target the
information about the process is contained in the averages
which depend neither on the direction nor on the distance
from the center of the nucleus (see Fig. \ref{fig-annih}).
This corresponds to the averaging over the outer region.
Assuming the annihilation cross section locally proportional 
to the density of nucleons (protons or neutrons) 
one sees that the ratio of the cross section of stopping the antiproton 
on neutrons to the cross section of stopping antiproton on
protons is proportional to 
corresponding ratio of the densities $\rho_N/\rho_Z$.
The total annihilation ratio is thus approximately proportional 
to the integral of the ratio of the densities over the whole 
outer region. A similar averaging has been performed in calculations 
of $\Delta_B$ 
and therefore the result given in Eq. (\ref{eq-nef}) i.e., NEF
is proportional to the halo factor $f_{\rm exp}$ (see Eq. (\ref{eq-hexp})) 
defined by\cite{Lubinski98,Schmidt99}
\begin{equation}
f_{\rm exp} \sim \frac{Z}{N} \frac{{\cal N}_{\bar pn}}{{\cal N}_{\bar pp}}\,,
\label{eq-hexp}
\end{equation}
where ${\cal N}_{\bar pp}$ and ${\cal N}_{\bar pn}$
denote the number of annihilations of both $\bar pp$ and $\bar pn$ 
types respectively (the final number of $_{Z-1}X^N$ nuclei
corresponding to $\bar pp$ annihilation
and the final number of $_ZX^{N-1}$ nuclei which correspond
to $\bar pn$ annihilation of $\bar p$ on a nucleus $_ZX^N$)
The final nuclei are "counted" after the anihilation processes 
has taken place.

\section{Results}
In the following section we compare our results of calculations of 
$\Delta_B$ to the experimentally measured halo factor $f_{\rm exp}$.

\begin{figure}[h]
\begin{center}
\includegraphics[scale=0.6]{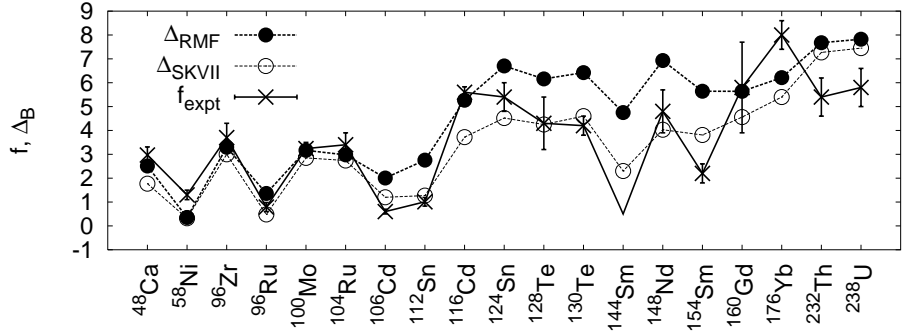}
\caption{Neutron excess factor $\Delta_B$ (NEF)
for nuclei studied by the Low Energy Antiprotonic Ring (LEAR) 
group at CERN. Experimental data are given with crosses connected 
by solid line.
Plain circles represent the RMF data and the open circles (dashed line)
correspond to Skyrme-Hartree calculations with SkVII force.
For more details and the references see the text. \label{fig-nef}}
\end{center}
\end{figure}

Figure \ref{fig-nef} shows the outer neutron excess factor ($\Delta_B$) 
as calculated  in the Relativistic Mean Field model (RMF, plain circles) 
with NL3 set of parameters\cite{LKR96} and in the selfconsistent 
Skyrme-Hartree (SH) model with SkVII force (open circles).\cite{skvii}
The data are presented for nuclei reported by Low Energy Antiprotonic 
Ring group (LEAR) at CERN.\cite{Lubinski98,Schmidt99}
It is observed that the calculations in both RMF and SH models
are strongly correlated with experimental halo factor $f_{\rm exp}$
(the scale of $\Delta_B$ is chosen arbitrarilly). The correlation shows
that the NEF $\Delta_B$ is a good indicator of the neutron halo property
and it gives the measure of the peripheral annihilation ratio 
of the antiproton on the neutrons and protons in the nuclear system.


\begin{thebibliography}{99}
\bibitem{Lubinski98} P. Lubi\'nski, J. Jastrz{\c e}bski, A. Trzci\'nska,
 W. Kurcewicz, F.J. Hartmann, R. Schmidt, T. von Egidy,
 R. Smola\'nczuk, and S. Wycech, Phys. Rev. C {\bf 57}, 2962, (1998) 
\bibitem{Schmidt99} R. Schmidt, F. J. Harrtmann, B. Ketzer, T. von Egidy, 
 T. Czosnyka, J. Jastrz\c ebski, M. Kisieli\'nski, P. Lubi\'nski, 
 L. Pie\'nkowski, A. Trzci\'nska, B. K\l os, R. Smola\'nczuk, 
 S. Wycech, W. P$\ddot{o}$schl, K. Gulda, W. Kurcewicz, and E. Widmann. 
 Phys. Rev. C {\bf 60}, {\bf 054309}, (1999) 
\bibitem{Baran97} A. Baran, K. Pomorski, and M. Warda,
Z. Phys A {\bf 357}, 33, (1997).
\bibitem{Baran04} A. Baran, P. Mierzyñski, submitted to Acta Physica 
Polonica B.
\bibitem{Bethe70}
H.~A. Bethe, {\em Ann. Rev. Nucl. Sci.}, {\bf 21}, 93, (1970).
\bibitem{LKR96}
G.~Lalazissis, J.~K{\"o}nig, , and P.~Ring.
Phys. Rev. C {\bf 55}, 540, (1996).
\bibitem{skvii} M. J. Giannoni, P. Quentin, Phys. Rev.,  {\bf 21}, 2076, (1980).
\end{thebibliography}
\end{document}